\documentclass[10pt,aps,reprint]{revtex4-1}
\usepackage{graphicx}
\usepackage{hyperref}

\makeatother
\begin{document}

\title{Comment on `Update of \textsuperscript{40}K and \textsuperscript{226}Ra and \textsuperscript{232}Th series $\gamma$-to-dose conversion factors for soil'}

\author{Alex Malins}\email{malins.alex@jaea.go.jp}
\affiliation{Center for Computational Science \& e-Systems, Japan Atomic Energy Agency, 178-4-4 Wakashiba, Kashiwa-shi,
Chiba-ken, 277-0871, Japan}

\author{Masahiko Machida}
\affiliation{Center for Computational Science \& e-Systems, Japan Atomic Energy Agency, 178-4-4 Wakashiba, Kashiwa-shi,
Chiba-ken, 277-0871, Japan}

\author{Kimiaki Saito}
\affiliation{Fukushima Environmental Safety Center, Sector of Fukushima Research and Development, Japan Atomic Energy Agency, 2-2-2, Uchisaiwai-cho, Chiyoda,
Tokyo, 100-8577, Japan}

\date{\today}

\begin{abstract}
A letter to the editor of the Journal of Environmental Radioactivity on the article: E. Gasser, A. Nachab, A. Nourreddine, Ch. Roy, and A. Sellam, `Update of \textsuperscript{40}K and \textsuperscript{226}Ra and \textsuperscript{232}Th series $\gamma$-to-dose conversion factors for soil', J.~Environ.~Radioactiv.~\textbf{138}, 68-71 (2014), DOI: \href{http://dx.doi.org/10.1016/j.jenvrad.2014.08.002}{10.1016/j.jenvrad.2014.08.002}.
\end{abstract}

\maketitle

Recently \citet{Gasser2014} have presented updated conversion factors between gamma radiation air kerma and concentrations of \textsuperscript{40}K and the \textsuperscript{226}Ra and \textsuperscript{232}Th series radionuclides within soil. Such conversion factors are important for calculating external exposures due to naturally occurring radioactive elements in the earth's crust~\citep{UNSCEAR2000}. The authors derived the conversion factors using MCNPX Monte Carlo simulations. The reported conversion factors are on average 20\% smaller than established estimates from both experiment~\citep{Beck1972} and simulation~\citep{Saito1995} (Table~5 of their paper). They attributed the difference as mainly due to their use of more up to date emission probabilities and branching ratios for radioactive decay. 

In our view the predominant cause of the difference is in fact a finite size issue with their simulations. Their simulation setup consisted of a cylindrical source containing radionuclides homogeneously within soil down to a depth of 1~m, and with varying radius up to 35~m. The air kerma was calculated for 1~m above the ground surface. From Fig.~3 in their paper they concluded that a 15~m radius source cylinder is sufficiently large for the conversion factors to reach their asymptotic values for a true half-space geometry. The infinite half-space is the applicable geometry for the conversion factors defined by convention in ICRU publication~53~\citep{ICRU53}.

\begin{figure}
\centering
\includegraphics[width=0.45\textwidth]{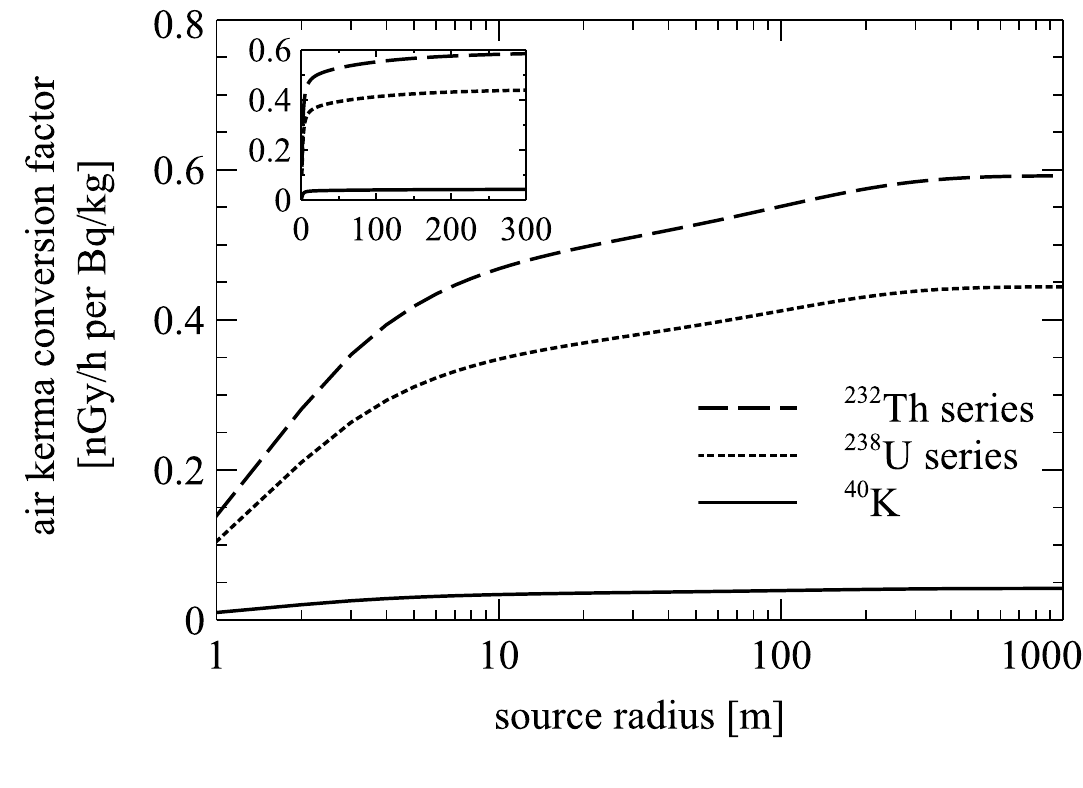}
\caption{Conversion factors for air kerma per unit concentration in soil with increasing source radius.}
\label{fig:kerma_source_radius}
\end{figure}

\begin{table*}
\centering
\begin{tabular}{ l l l l l l }\hline
 Calculation & Source radius & \multicolumn{3}{l}{Conversion factor (nGy/h per Bq/kg)} & Nuclear data ref.\\ \cline{3-5}
  &  (m) & \textsuperscript{40}K & \textsuperscript{238}U series &  \textsuperscript{232}Th series & \\ \hline
\citet{Gasser2014} & 35 &	0.036	& 0.357%\textsuperscript{\hyperlink{label}{a}}
\textsuperscript{a}	& 0.482 & \citet{Nucleide-Lara2013} \\
This letter & 35 &	0.037	& 0.383	& 0.515 & \citet{ICRP107} \\ 
\citet{Saito1995} & $\infty$ &	0.042	& 0.463	& 0.604 & \citet{JEF-2.2} \\
This letter & $\infty$ &	0.042	& 0.444	& 0.592 & \citet{ICRP107} \\\hline
\end{tabular}
\caption{Summary of different calculations for air kerma conversion coefficients. %\textsuperscript{\hypertarget{label}{a}}
\textsuperscript{a}Calculation for \textsuperscript{226}Ra \& progeny only.}
\label{tab:conv_factors}
\end{table*}

A 35~m radius simulation is not sufficient to model a half-space, as can be seen from Fig.~\ref{fig:kerma_source_radius} in this letter. The figure displays some of our results covering a much larger range of source radii than considered by Gasser~\textit{et~al.} Our results were calculated using the PHITS radiation transport code~\citep{Sato2013}. We considered source cylinders with 1~m depth in soil, and radii up to 1000~m. The natural radionuclides were distributed homogeneously within the cylinder, and the air and soil compositions and densities followed \citet{Eckerman1993}. The photon emission probabilities and decay branching fractions were drawn the recent ICRP publication~107 on nuclear decay data~\citep{ICRP107}.

The conversion factors increase monotonically in the range 1 to 35~m as studied by \citet{Gasser2014}. However, they continue to increase beyond 35~m radius source cylinders, and only attain their limiting values to three significant figures for source cylinders greater than 700~m in radius. The apparent points of inflection on the curves in our Fig.~\ref{fig:kerma_source_radius} are artifacts of the logarithmic abscissa, as can be seen from the inset which shows the same data on a linear scale.

Our results yield air kerma conversion coefficients of 0.042, 0.444 and 0.592 (nGy/h per Bq/kg) for \textsuperscript{40}K, the \textsuperscript{238}U series and the \textsuperscript{232}Th series, respectively, in the asymptotic limit. These values are reasonably consistent with the values of 0.042, 0.463 and 0.604 (nGy/h per Bq/kg) respectively found in~\citet{Saito1995} (see Table~\ref{tab:conv_factors} of this letter).  Similarly, results recently published by \citet{Askri2015} from simulations of large radius source cylinders ($\simeq 500$~m) were consistent with \citet{Saito1995}. However, the results contrast with those of \citet{Gasser2014}, namely 0.036, 0.357 and 0.482 (nGy/h per Bq/kg) respectively for \textsuperscript{40}K, the \textsuperscript{226}Ra series and the \textsuperscript{232}Th series.

For direct comparison we simulated source cylinders with radius 35~m. In this case the conversion factors obtained were 0.037, 0.383 and 0.515 (nGy/h per Bq/kg) for \textsuperscript{40}K, the \textsuperscript{238}U series and the \textsuperscript{232}Th series, respectively. These values are in reasonable agreement with those of~\citet{Gasser2014}. The residual differences between the two estimates for 35~m radius sources, and between our latest estimates and \citet{Saito1995} for the true half-space geometry, are of the order of a few percent. Therefore most of the $\simeq 20$\% reduction in the conversion factors obtained by \citet{Gasser2014} can be explained by a finite size effect. Possible explanations for the residual differences once this is accounted for are different sources for decay emission data, as suggested by \citet{Gasser2014}, or other detail variations between the calculation methods. These could be differences in the energy resolution of the Monte Carlo simulations, statistical accuracy, or the method used for interpolating between the photon fluence rate to air kerma conversion coefficients, for instance.

We encourage Gasser~\textit{et~al.}~to increase the size of their simulations and recalculate the air kerma conversion coefficients. Larger radius source cylinders can be simulated quickly and efficiently by transforming the Monte Carlo source and detector volumes as outlined by~\citet{Namito2012}. Another cheap option is to use a plane detector and periodic boundary conditions. Modeling a source with a sufficiently large radius will yield results closer to the conversion factors established in the literature. Any remaining differences between their results and literature values may then be attributed to the use of more up-to-date decay data with lower emission probabilities, and perhaps other methodological differences as discussed above.

\bibliography{joer_letter_aam}

\end{document}